\documentclass[showkeys,showpacs,superscriptaddress]{revtex4-2}
\usepackage{amsmath}
\usepackage{amssymb}
\usepackage{graphicx}
\usepackage{float}
\usepackage{xcolor}
\usepackage{physics}
\usepackage{ulem}
\allowdisplaybreaks

\begin{document}

\title{
Proca balls with angular momentum or flux of electric field
}

\author{
Vladimir Dzhunushaliev
}
\email{v.dzhunushaliev@gmail.com}
\affiliation{
Department of Theoretical and Nuclear Physics,  Al-Farabi Kazakh National University, Almaty 050040, Kazakhstan
}
\affiliation{
Institute of Nuclear Physics, Almaty 050032, Kazakhstan
}
\affiliation{
Academician J.~Jeenbaev Institute of Physics of the NAS of the Kyrgyz Republic, 265 a, Chui Street, Bishkek 720071, Kyrgyzstan
}

\author{Vladimir Folomeev}
\email{vfolomeev@mail.ru}
\affiliation{
Institute of Nuclear Physics, Almaty 050032, Kazakhstan
}
\affiliation{
Academician J.~Jeenbaev Institute of Physics of the NAS of the Kyrgyz Republic, 265 a, Chui Street, Bishkek 720071, Kyrgyzstan
}
\affiliation{
Laboratory for Theoretical Cosmology, International Centre of Gravity and Cosmos,
Tomsk State University of Control Systems and Radioelectronics (TUSUR),
Tomsk 634050, Russia
}


\begin{abstract}
Within SU(2) Higgs-Proca theory, we obtain a family of nontopological static solutions describing localized,
finite-energy configurations (Proca balls). The gauge symmetry of the theory is explicitly broken by introducing
a vector Proca field whose components have different masses. Such solutions describe particlelike systems,
the crucial feature of which is that they either possess a nonzero total angular momentum or have a flux of electric field through the plane of symmetry of such objects.
It is shown that the angular momentum is provided by static crossed electric and magnetic fields.
The existence of the solutions is caused by the fact that we circumvent the conditions of the no-go theorem,
according to which there are no stationary and axially symmetric spinning excitations for the 't~Hooft-Polyakov monopoles, Julia-Zee dyons, sphalerons, and also vortices.
The dependence of some integral physical quantities on the ratio of the Proca-field masses is studied. It is demonstrated that the inclusion of external sources (charges) enables
one to obtain solutions with equal Proca-field masses. We also discuss the possibilities of using quarks as sources of the Proca field under investigation and for treating
the Proca balls as glueballs in SU(2) Higgs-Proca theory.
\end{abstract}


\date{\today}

\maketitle

\section{Introduction}

There are numerous studies in the literature on solitonlike solutions to classical nonlinear equations for various fundamental fields. The important property of such solutions is that they describe finite-energy objects with a localized, nondispersive energy density. Most of the solitonlike solutions have an essentially nonperturbative nature, i.e., they cannot be obtained by starting from solutions of the
corresponding linear part of the field equations and treating the nonlinear terms perturbatively.
The presence of the nonlinearity compensates the field's natural tendency to disperse.
A role of such nonlinearities in the system may be played by self-interaction of the fields or by any other fields (for example, gravitational ones).

In this connection, the search for solitonlike solutions possessing some new properties is an interesting and rather complicated problem. For instance, the question can be asked whether solitonlike solutions with a nonzero total angular momentum do exist? The answer to this question is positive. For example,
a solution with a nonvanishing angular momentum describing localized configurations ({\it Q}-balls) in a theory with a nonlinear scalar field was obtained in Ref.~\cite{Volkov:2002aj}.
In Ref.~\cite{Kleihaus:2005me}, this solution has been generalized to the case of the presence of a gravitational field. On the other hand, in Ref.~\cite{Volkov:2003ew},
it was shown that in  SU(2) Yang-Mills-Higgs gauge theory there are no stationary and axially symmetric spinning excitations for all known topological solitons in the one-soliton sector;
that is, for the 't~Hooft-Polyakov monopoles, Julia-Zee dyons, sphalerons, and also vortices. Nevertheless,
it was demonstrated in Refs.~\cite{Radu:2008ta,Kleihaus:2008cv} that it is possible to circumvent the conditions imposed by this no-go theorem by
considering $\text{SU(2)}\times \text{U(1)}$ theory involving complex Higgs scalar fields and a U(1) gauge field; this enables one to obtain solutions
possessing a finite energy and nonzero angular momentum.
A similar problem occurs in studying the structure of the proton spin when it is assumed that gluons may also contribute to the proton spin~\cite{Shuryak2021}.

In this study, we show that within SU(2) Higgs-Proca theory, where the gauge symmetry is explicitly broken by introducing a massive vector (Proca) field,
localized regular solutions describing configurations with a nonvanishing total angular momentum can exist. In what follows we will refer to such configurations as Proca balls (or, for brevity, {\it P}-balls).
Interest in studying Proca fields in various aspects  has increased considerably in recent years because of the possibility of obtaining new solutions
suitable for the description of various physical objects and processes. In particular, these can be compact, strongly gravitating
starlike  configurations~\cite{Brito:2015pxa,Herdeiro:2017fhv,Dzhunushaliev:2019kiy,Herdeiro:2019mbz,Dzhunushaliev:2019uft,Bustillo:2020syj},
black holes~\cite{Heisenberg:2017xda,Kase:2018voo}, dark matter~\cite{Arkani-Hamed:2008hhe,Pospelov:2008jd}, processes at cosmological scales~\cite{DeFelice:2016yws,DeFelice:2016uil},
various effects related to the possible presence of the rest mass of a photon~\cite{Tu:2005ge}, as well as flux tube configurations filled with electric and
magnetic fields~\cite{Dzhunushaliev:2019sxk,Dzhunushaliev:2020eqa,Dzhunushaliev:2021uit}.

As another interesting problem, one may consider the question of the existence of solitonlike solutions describing localized configurations (for example, tubes) with a nonzero flux of electric field through the plane of symmetry of such systems.  A study of such configurations may be of considerable interest
from the point of view of the confinement problem in quantum chromodynamics where the existence of a flux of electric field confined inside a tube connecting quarks and antiquarks is necessary~\cite{Shuryak2021}.

In our recent paper~\cite{Dzhunushaliev:2021oad}, we have shown that in SU(2) Higgs-Proca theory containing one real Higgs scalar field coupled to Proca fields there exist localized flux tube solutions with a flux of magnetic field through the plane of symmetry of a tube. In turn, in the presence of a gravitational field,  such systems possess an axially symmetric dipole field (a Proca dipole) sourced by a current associated with the Higgs field~\cite{Dzhunushaliev:2021vwn}. In the present paper, we consider particlelike, topologically trivial solutions in SU(2) Higgs-Proca theory where the gauge symmetry is explicitly broken and which contains a triplet of real Higgs scalar fields.  Our purpose will be to obtain regular localized solutions describing {\it P}-ball type objects possessing such physically interesting properties like the presence of either a nonvanishing total angular momentum or a flux of electric field through the plane of symmetry of such objects.

The paper is organized as follows. In Sec.~\ref{Higgs-Proca_theory}, we write down the general field equations for SU(2) Higgs-Proca theory, which we use in Sec.~\ref{ansatz_eqs} to derive the equations for a suitable field ansatz. In Sec.~\ref{inf_tube}, we obtain cylindrically symmetric solutions to the equations of Sec.~\ref{ansatz_eqs} describing infinite tubes with a flux of longitudinal color magnetic field and nonzero linear angular momentum density along the tube axis. In Sec.~\ref{fin_tube}, we find axially symmetric solutions to the equations of Sec.~\ref{ansatz_eqs} (both with and without charge and current densities) describing finite-size tubes either with a nonzero total angular momentum~(Sec.~\ref{non_zero_ang_mom}) or with a flux of longitudinal chromoelectric field~(Sec.~\ref{el_flux}). Finally, in Sec.~\ref{concl}, we summarize and discuss the results obtained in the present paper.

\section{SU(2) Higgs-Proca theory}
\label{Higgs-Proca_theory}

The Lagrangian describing a system consisting of a non-Abelian SU(2) Proca field $A^a_\mu$ coupled to a triplet of real Higgs scalar fields $\phi^a$ can be taken in the form (hereafter, we work in units such that $c=\hbar=1$)
\begin{equation}
	\mathcal L =  - \frac{1}{4} F^a_{\mu \nu} F^{a \mu \nu} -
	\frac{1}{2}\left( \mu^2 \right)^{a b, \mu}_{\phantom{a b,}\nu}
	A^a_\mu A^{b \nu} +
	\frac{1}{2} D_\mu \phi^a D^\mu \phi^a  -
	\frac{\Lambda}{4} \left( \phi^a \phi^a - v^2 \right)^2.
\label{0_10}
\end{equation}
Here
$
	F^a_{\mu \nu} = \partial_\mu A^a_\nu - \partial_\nu A^a_\mu +
	g \epsilon_{a b c} A^b_\mu A^c_\nu
$ is the field strength tensor for the Proca field, where $\epsilon_{a b c}$ are the SU(2) structure constants (the completely antisymmetric Levi-Civita symbol), $g$ is the coupling constant,
$a, b, c = 1, 2, 3$ are SU(2) color indices, $\mu, \nu = 0, 1, 2, 3$ are spacetime indices; $D_\mu \phi^a = \partial_\mu \phi^a + g \epsilon^{a b c} A_\mu^b \phi^c$. The Lagrangian \eqref{0_10} also contains the arbitrary constants $v$ and $\Lambda$ and the Proca field mass tensor
$
	\left( \mu^2 \right)^{a b, \mu}_{\phantom{a b,}\nu}
$, which we suppose to be symmetric with respect to the color and spacetime indices.

Using Eq.~\eqref{0_10}, the corresponding field equations can be written in the form
\begin{align}
	D_\nu F^{a \mu \nu} +
	 \left( \mu^2 \right)^{a b, \mu}_{\phantom{a b,}\nu} A^{b \nu}
	=& g \epsilon^{abc} \phi^b D^\mu \phi^c + j^{a \mu},
\label{0_20}\\
	D_\mu D^\mu \phi^a =&
	- \Lambda \phi^a \left(
		\phi^b \phi^b - v^2
	\right),
\label{0_30}
\end{align}
where, for the sake of generality, we have also added the current four-vector $ j^{a \mu}$.
For such a system, the symmetric energy-momentum tensor (EMT) coming from the Lagrangian~\eqref{0_10} is
\begin{equation}
\begin{split}
	T_{\mu\nu} & = - F^a_{\mu\alpha}F^{a \alpha}_\nu
	+ \frac{1}{4}g_{\mu\nu} F^{a}_{\alpha\beta} F^{a \alpha\beta}
	+ D_\mu\phi^a D_\nu\phi^a - g_{\mu\nu} \left[
		\frac{1}{2}D_\alpha\phi^a D^\alpha\phi^a
		- \frac{\Lambda}{4} \left(\phi^a \phi^a -v^2\right)^2
	\right]
\\
	& - \left( \mu^2 \right)^{a b, \alpha}_{\phantom{a b,} \nu}A^a_\alpha A^{b}_\mu
	+ \frac{1}{2} g_{\mu\nu}\left( \mu^2 \right)^{a b, \alpha}_{\phantom{a b,}\beta}
	A^a_\alpha A^{b \beta}.
\end{split}
\label{emt_gen}
\end{equation}

Here, the following remark is to be made. The above EMT was obtained by varying the Lagrangian~\eqref{0_10} with respect to the metric. However, the Lagrangian contains the Proca field mass tensor with mixed spacetime indices which, in principle, can be raised/lowered by introducing the corresponding metric tensor. In obtaining the EMT~\eqref{emt_gen}, we did not do it, but only lowered the spacetime index on the vector potential $A^{b \nu}$, assuming that the Proca field mass tensor
with mixed spacetime indices is regarded as a fundamental tensor. However, if we would drop this assumption and suppose that the Proca field mass tensor with contravariant or covariant indices is fundamental, then, before performing the variation, it would be necessary to raise/lower the spacetime index on the Proca field mass tensor in Eq.~\eqref{0_10}. As a result, there would already be two metric tensors, with respect to which the variation would have to be carried out. Then the resulting EMT would contain a mass term different from that given in Eq.~\eqref{emt_gen}. It is evident that such an inherent ambiguity in defining the EMT is related to our initial assumption that different components of the vector potential of the Proca field have different masses. As will be demonstrated below, in considering a physically more realistic situation where the system contains charges (for example, quarks), it is possible to obtain solutions with equal masses of all components of the vector potential (i.e., the Proca field mass tensor becomes a scalar quantity).

Making use of Eq.~\eqref{emt_gen},  the field energy density of the system under consideration can be recast in the form
\begin{equation}
\begin{split}
	\varepsilon = &\frac{1}{2} \left( E^a_i \right)^2 +
	\frac{1}{2} \left( H^a_i \right)^2 -
			\left( \mu^2 \right)^{a b, \alpha}_{\phantom{a b,} 0} A^a_\alpha A^b_0 +
		\frac{1}{2} \left( \mu^2 \right)^{a b, \alpha}_{\phantom{a b,} \beta} A^a_\alpha A^{b \beta}
		+
	\frac{1}{2} \left( D_t \phi^a \right)^2 +
	\frac{1}{2} \left( D_i \phi^a \right)^2
+	\frac{\Lambda}{4} \left( \phi^a\phi^a - v^2 \right)^2 ,
\label{0_40}
\end{split}
\end{equation}
where $i=1,2,3$ and $E^a_i$ and $H^a_i$ are the components of the electric and magnetic field strengths, respectively.

\section{The ansatz and equations
}
\label{ansatz_eqs}

Our purpose is to obtain regular localized axially symmetric solutions in SU(2) Higgs-Proca theory.
Consistent with this, we take the ansatz for the Proca and scalar fields  in the form
\begin{equation}
	A^1_t = \frac{ f(\rho, z)}{g} , \;
	A^1_\varphi =  \frac{\rho \, k(\rho, z)}{g} , \;
	A^3_t =  \frac{ h(\rho, z)}{g} , \;
	A^3_\varphi = \frac{\rho \, w(\rho, z)}{g} , \;
 	\phi = \left\{\phi_1(\rho, z), 0,  \phi_3(\rho, z)\right\}
\label{1_10}
\end{equation}
written in cylindrical coordinates  $\{t, \rho, \varphi, z\}$. For such a choice, there are the following
nonvanishing color electric and magnetic fields (physical components),
\begin{align}
	E^1_z = & - \frac{ f_{, z}}{g} , \quad
	E^3_z = - \frac{ h_{, z}}{g}, \quad
	E^1_\rho = - \frac{f_{, \rho}}{g} , \quad
	E^3_\rho = - \frac{h_{, \rho}}{g},
\label{fields_5}\\
	H^1_z = & - \frac{
		 \rho \, k_{, \rho} +  k
	}{g \rho} , \quad
	H^3_z = - \frac{
		 \rho \, w_{, \rho} +  w
	}{g\rho} , \quad
	H^1_\rho = \frac{ k_{, z}}{g} ,  \quad
	H^3_\rho = \frac{ w_{, z}}{g} ,
\label{fields_10}
\end{align}
where a comma in lower indices denotes differentiation with respect to the corresponding coordinate.

In what follows we will consider the simplest case where
\begin{equation}
	h = - f, k = -w, \phi_1 = \phi_3 = \phi.
\label{condition}
\end{equation}
In this case Eqs.~\eqref{0_20} and \eqref{0_30}, after substitution of the components~\eqref{1_10}, yield
\begin{align}
	\Delta_{x,y} f  - 2 \phi^2 f+ f = & j^{1 t},
\label{1_20}\\
	\Delta_{x,y} w - \frac{w}{x^2} -2  \phi^2 w+ \alpha^2 w = & j^{3\varphi},
\label{1_30}\\
	\Delta_{x,y} \phi
	+  \left[
		2 \left( f^2 - w^2\right)  - \Lambda \left( 2 \phi^2 - v^2\right)
	\right]\phi = & 0 ,
\label{1_40}
\end{align}
where $\Delta_{x,y}= \partial_{xx} + \partial_{yy}  + \partial_x / x$ is the Laplacian in the coordinates $x,y$.
Equations~\eqref{1_20}-\eqref{1_40} are written in terms of the dimensionless variables
$$
	\left( x, y\right) = \mu_f \left( \rho, z\right) , \quad
	\left(\tilde \phi, \tilde v\right) = \frac{g}{\mu_f} \left( \phi,  v\right), \quad
	\left( \tilde f,  \tilde w\right) = \frac{1}{\mu_f} \left(f,  w \right) , \quad
    \left( \tilde j^{1 t},  \tilde j^{3\varphi}\right) = \frac{1}{\mu_f^3} \left( j^{1 t},  j^{3\varphi} \right) , \quad
	\tilde\Lambda = \frac{ \Lambda}{g^2}, \quad
	\alpha = \frac{\mu_w}{\mu_f},
$$
with the following components of the Proca field mass tensor:
$\mu_{f}^2\equiv\left( \mu^2 \right)^{1 1, t}_{\phantom{a b,}t}$ and
$\mu_{w}^2\equiv\left( \mu^2 \right)^{3 3, \varphi}_{\phantom{a b,}\varphi}$. To make the notation simpler, we have omitted
the tilde sign over the dimensionless variables in Eqs.~\eqref{1_20}-\eqref{1_40}. Also notice that in the absence of the currents, Eqs.~\eqref{1_20} and \eqref{1_30} can be regarded as Schr\"{o}dinger-type equations with the potentials  $2 \phi^2$ and
$\left(1/x^2+2 \phi^2\right)$, respectively. This enables us to assume that regular solutions to these equations can exist only if a solution for the scalar field $\phi$ ensures the existence of a potential well which allows the presence of positive energy levels; this, in turn, assumes that $\phi \rightarrow v/\sqrt{2}$ at infinity.

We must emphasize here the following point; although Eqs.~\eqref{1_20} and \eqref{1_30} look like equations for U(1) Proca scalar electrodynamics, this is not so, however.
The reason is that if we ignore the simplifying assumption~\eqref{condition}, an equation, for example, for the potential $f$ would take the form
\begin{equation}
	\Delta_{\rho,z} f -  \left( w^2 + g^2 \phi_3^2\right) f + h k w + g^2 h \phi_1 \phi_3 + \mu^2_f f =  j^{1 t} .
\label{nonAbelian}
\end{equation}
This equation already contains the nonlinear terms  $f w^2$ and $h k w$, and this reflects the non-Abelianity of the theory under consideration. A similar situation occurs for other equations as well.

Finally, the Poynting vector is
\begin{equation}
	S^i = \frac{\epsilon^{i j k}}{\sqrt{\gamma}} E^a_j H^a_k ,
\end{equation}
where  $\gamma$ is the determinant of the space metric. For the field strengths~\eqref{fields_5} and \eqref{fields_10}, this expression gives the following nonvanishing dimensionless physical component,
\begin{equation}
	\tilde 	S_\varphi\equiv \frac{ g^2}{\mu_f^4} S_\varphi =
	2\left[f_{, x} \left(w_{, x } + \frac{w}{x}\right)  +  f_{, y} w_{, y}\right].
\label{UmPoynt}
\end{equation}

\section{Infinite flux tube solutions}
\label{inf_tube}

Before proceeding to finding localized solutions (that is, solutions possessing a finite total energy and linear sizes), we will obtain solutions describing an infinite tube filled with electric and magnetic fields. For such a tube, the derivatives with respect to $y$ in Eqs.~\eqref{1_20}-\eqref{1_40} vanish, and we have the following set of equations (without charges and currents),
\begin{align}
	f^{\prime\prime}+\frac{1}{x}f^\prime - 2 \phi^2 f+ f = & 0,
\label{1_20inf}\\
	w^{\prime\prime}+\frac{1}{x}w^\prime - \frac{w}{x^2} -2  \phi^2 w+ \alpha^2 w = & 0,
\label{1_30inf}\\
	\phi^{\prime\prime}+\frac{1}{x}\phi^\prime
	+  \left[
		2 \left( f^2 - w^2\right)  - \Lambda \left( 2 \phi^2 - v^2\right)
	\right]\phi = & 0 ,
\label{1_40inf}
\end{align}
where the prime denotes differentiation with respect to $x$. Such a tube contains the following nonvanishing color electric and magnetic fields (physical components) obtained from Eqs.~\eqref{fields_5} and~\eqref{fields_10}
\begin{equation}
 E^1_x =- E^3_x=  -f^\prime ,  \quad
 H^1_y=- H^3_y =   \frac{x w^\prime + w }{ x },
\label{elec_magn_str}
\end{equation}
where the dimensionless $\left(E^1_x, H^1_y\right)=\left(g/\mu_f^2\right)\left(E^1_\rho, H^1_z\right)$.

We seek a solution to Eqs.~\eqref{1_20inf}-\eqref{1_40inf} in the vicinity of the origin of coordinates in the form
\begin{eqnarray}
	 f(x) &=& f_0 +  f_2 \frac{x^2}{2} + \dots \quad \text{with} \quad
 f_2 = \frac{f_0}{2} \left(
		 2 \phi_0^2 -  1	\right) ,
\label{3_a_100}\\	
 w(x) &=&  w_1 x + \dots ,
\label{3_a_115}\\
	\phi(x) &=&  \phi_0 +  \phi_2 \frac{x^2}{2} + \dots \quad \text{with} \quad
 \phi_2 =  \phi_0 \left[
		 \Lambda\left( \phi_0^2 -\frac{1}{2}v^2\right)-f_0^2
		\right],
\label{3_a_120}
\end{eqnarray}
where the expansion coefficients $f_0,  \phi_0$, and $w_1$ are arbitrary.

In turn, the asymptotic behavior of the functions $f,  w$, and $\phi$ can be found from Eqs.~\eqref{1_20inf}-\eqref{1_40inf} in the form
\begin{equation}
	 f(x) \approx  f_{\infty}
	\frac{e^{- x \sqrt{ v^2 -  1}}}{\sqrt x}, \quad
	w(x) \approx  w_{\infty}
	\frac{e^{- x \sqrt{ v^2 -  \alpha^2}}}{\sqrt x},\quad
	 \phi(x) \approx  \frac{v}{\sqrt{2}} -  \phi_\infty
	\frac{e^{- x \sqrt{2 \Lambda  v^2}}}{\sqrt x} ,
\label{3_a_160}
\end{equation}
where $f_{\infty},   w_{\infty}$, and $\phi_\infty$ are integration constants.

\begin{figure}[H]
\includegraphics[width=1.\linewidth]{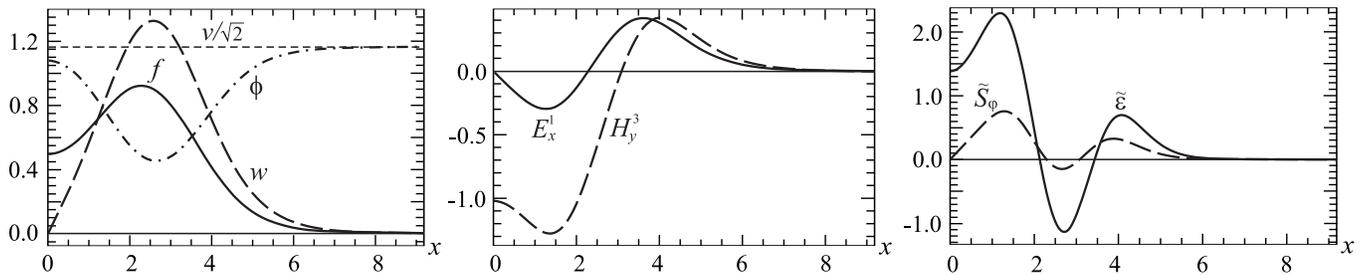}
\vspace{-0.3cm}
\caption{Infinite flux tube solutions with the system parameters $v = 1.65,	\Lambda = 0.4,  \alpha=1.1$. Left panel shows the profiles of the Proca field potentials $f $ and $w$ and of the scalar field $\phi$; middle panel~-- the profiles of the color electric, $ E^1_x$, and magnetic, $H^3_y$, fields	from Eq.~\eqref{elec_magn_str}; right panel~-- the profiles of the linear energy density from Eq.~\eqref{inf_energ_dens} and of the linear momentum density given by Eq.~\eqref{UmPoynt_inf}.
}
\label{fig_inf_tube}
\end{figure}

The derivation of solutions to the set of equations~\eqref{1_20inf}-\eqref{1_40inf} is an eigenvalue problem for the system parameters $\alpha, \Lambda$, and $ v$ (or, equivalently, for the expansion coefficients $f_0,  \phi_0$, and $w_1$). The numerical solution describing the behavior of the Proca field potentials and of the corresponding electric and magnetic fields is given in Fig.~\ref{fig_inf_tube}. This figure also shows the energy density obtained from Eq.~\eqref{0_40} using~\eqref{1_10} and~\eqref{elec_magn_str} in the form
\begin{equation}
	\tilde\varepsilon \equiv \frac{g^2}{\mu_f^4}\varepsilon =
f_{, x}^2+w_{, x}^2+\phi_{, x}^2+2\frac{w w_{, x}}{x}+2\left(f^2+ w^2\right)\phi^2
- f^2-\alpha^2 w^2+\frac{w^2}{x^2}+\frac{\Lambda}{4}\left(2\phi^2-v^2\right)^2 .
\label{inf_energ_dens}
\end{equation}

It is seen from Fig.~\ref{fig_inf_tube} that the cylindrically symmetric solutions obtained can be used for the description of infinitely long tubes within non-Abelian Higgs-Proca theory under consideration. Such tubes possess the finite linear energy density,
\begin{equation}
	\mathcal{E} = 2 \pi \int \limits_0^\infty \varepsilon(\rho) \rho  d \rho
	< \infty ,
\end{equation}
and the finite flux of the longitudinal color magnetic field $H^1_z$,
\begin{equation}
	\Phi_z^H = 2 \pi \int \limits_0^\infty  H^1_z \rho d \rho < \infty .
\end{equation}
A similar flux occurs for the field $H^3_z$ in a direction opposite to that of for the component $H^1_z$.

Also, the system contains the following nonvanishing component of the energy flux/momentum density [cf. Eq.~\eqref{UmPoynt}]
\begin{equation}
\tilde 	S_\varphi= 2 f_{, x} \left(w_{, x } + \frac{w}{x}\right),
\label{UmPoynt_inf}
\end{equation}
whose behavior is demonstrated in the right panel of Fig.~\ref{fig_inf_tube}. The presence of such component ensures the presence in the system of a nonzero linear angular momentum density along the tube axis:	$\mathcal{L}_z = \rho S_\varphi$.

Note that, unlike the configurations considered by us earlier in Refs.~\cite{Dzhunushaliev:2019sxk,Dzhunushaliev:2020eqa,Dzhunushaliev:2021uit},
these tubes contain no flux of the longitudinal color electric field $E^a_z$.

\section{Finite flux tube solutions}
\label{fin_tube}

Having considered the infinite tube solutions, we turn now to the study of regular solutions with finite sizes/energies describing particlelike objects (Proca balls) possessing either a nonzero total angular momentum or a flux of electric field through the plane $z=0$. In both cases, the solutions will be non-Abelian ones, since, as pointed out above, regardless of the fact that the potentials
$A^1_{t, \varphi}$ and $A^3_{t, \varphi}$ belong to two subgroups
$\text{U(1)} \subset \text{SU(2)}$, the presence of the terms $f w^2$ and $h k w$ in Eq.~\eqref{nonAbelian} reflects the non-Abelianity of the system. The same remarks are applicable to the remaining equations as well.

\subsection{Boundary conditions}

In the present paper we choose such boundary conditions that the components of the electric field strength are even or odd functions of $z$. Consider first the case where the component $E^1_z$ is an odd function and  $E^1_\rho$ is an even function. This presupposes the use of the following boundary conditions
\begin{equation}
\begin{split}
	&\left. \frac{\partial f}{\partial \rho}\right|_{\rho = 0} =
	\left. \frac{\partial \phi}{\partial \rho}\right|_{\rho = 0} =  0,  \left. w \right|_{\rho = 0} = 0;
	\quad\left. \frac{\partial f}{\partial z}\right|_{z = 0} =
	\left. \frac{\partial w}{\partial z}\right|_{z = 0} =
	\left. \frac{\partial \phi}{\partial z}\right|_{z = 0} =  0; \\
	& f = w = 0,  \phi = \frac{v}{\sqrt{2}} \quad \text{ as } \quad \rho^2 +z^2 \to \infty .
\end{split}
\label{3_A_10}
\end{equation}
In what follows we will refer to configurations with such choice of the boundary conditions as the type A systems.

In turn, for the case where $E^1_z$ is an even function of $z$ and  $E^1_\rho$ is an odd function of $z$, we choose the boundary conditions
\begin{equation}
\begin{split}
	&\left. \frac{\partial f}{\partial \rho}\right|_{\rho = 0} =
	\left. \frac{\partial \phi}{\partial \rho}\right|_{\rho = 0} =  0,  \left. w \right|_{\rho = 0} = 0;
	\quad	\left. \frac{\partial w}{\partial z}\right|_{z = 0} =
	\left. \frac{\partial \phi}{\partial z}\right|_{z = 0} =  0,  \left. f \right|_{z = 0} = 0; \\
	& f = w = 0,  \phi = \frac{v}{\sqrt{2}} \quad \text{ as } \quad \rho^2 +z^2 \to \infty ,
\end{split}
\label{3_A_20}
\end{equation}
and refer to configurations with this choice as the type B systems.

\subsection{Asymptotic behavior }

Even before obtaining numerical solutions, it is possible to estimate their asymptotic behavior, keeping in mind that for the functions $f$ and $w$ we will seek solutions that decay exponentially with distance, and for the function $\phi$~-- a solution that goes exponentially to a constant. For this purpose, it is convenient to take a spherical coordinate system $\{r, \theta, \varphi\}$. In this case, as $r \rightarrow \infty$, the scalar field $\phi \approx v/\sqrt{2}-\eta \to v/\sqrt{2}$, and the functions $f, w,\eta\to 0$ exponentially fast. As a result, from Eqs.~\eqref{1_20}-\eqref{1_40}, one can obtain asymptotic equations
\begin{align}
	\bigtriangleup_{r, \theta} f -
	 \left(v^2-1\right) f= & 0 ,
\label{2_90}\\
  \bigtriangleup_{r, \theta} w - \frac{w}{r^2\sin^2{\theta}}  -
	\left(v^2- \alpha^2\right) w =& 0 ,
\label{2_100}\\
	  \bigtriangleup_{r, \theta} \eta -
	2  \Lambda v^2 \eta =& 0 ,
\label{2_110}
\end{align}
where $\bigtriangleup_{r, \theta}$ is the Laplacian in the coordinates $r,\theta$. Eqs.~\eqref{2_90} and \eqref{2_110} have obvious solutions in the form
\begin{align}
	f \approx & C_{f}
	\left( Y\right)^0_{l_f}
	\frac{e^{- r \sqrt{ v^2 - 1}}}{r},
\label{2_120}\\
	\eta \approx & C_{\eta}
	\left( Y\right)^0_{l_\eta}
	\frac{e^{- r \sqrt{2 \Lambda v^2}}}{r},
\label{2_130}
\end{align}
where $\left( Y\right)^0_{l_{f,\eta}}$ are spherical functions and $C_{f, \eta}$ are constants. In turn, Eq.~\eqref{2_100} has a solution similar to \eqref{2_120}, but only with the angular part expressed in terms of special functions (we do not show this expression here to avoid overburdening the text). The general solution of Eqs.~\eqref{2_90}-\eqref{2_110} represents a superposition of the above solutions, and it is obtained by summing over $l_f$ and $l_\eta$. It follows from the above expressions that there are lower limits for the parameter $v$ ensuring the exponential asymptotic decay of the solutions: $v>1$ and $v>\alpha$.

\subsection{Charges and currents}

When the system contains any charged particles (for example, quarks), the right-hand sides of the Proca equations \eqref{1_20} and \eqref{1_30} should contain nonvanishing current densities. In considering a self-consistent problem with a tube connecting quarks, such currents must be created by spinor fields describing quarks. For the sake of simplicity, here we consider a  toy model where the currents are given by hand. This means that the location of the quarks and the magnitude of color currents and charges created by them are fixed.

In the simplest case the currents can be given, for instance, by the Gaussian distribution. Bearing in mind the symmetry of the problems at hand, we will use the following expressions
\begin{align}
	&\text{For the type A systems}: \quad  j^{1 t} =
	\left( j_0\right)^{1 t} \exp\left(-\frac{x^2 + (y - l)^2}{R^2_0}\right), \quad
	 j^{3\varphi} = \left( j_0\right)^{3\varphi} \exp\left(-\frac{y^2 + (x - l)^2}{R^2_0}\right);
\label{1_300}\\
	&\text{For the type B systems}: \quad  j^{1 t} =
	\left( j_0\right)^{1 t} y \exp\left(-\frac{x^2 + (y - l)^2}{R^2_0}\right), \quad
	 j^{3\varphi} = \left( j_0\right)^{3\varphi} \exp\left(-\frac{y^2 + (x - l)^2}{R^2_0}\right),
\label{1_310}
\end{align}
where $\left( j_0\right)^{1 t},\left( j_0\right)^{3\varphi}, l$, and $R_0$ are arbitrary constants. Such a choice for $j^{1 t}$ implies that  a ``quark'' is located on the tube axis at a distance  $l$ from the origin of coordinates. In turn, the expression for $j^{3\varphi}$  implies the presence of the current of quarks in the $z=0$ plane located on a circle of radius $l$.

\subsection{Numerical approach}

The set of three coupled nonlinear elliptic partial differential equations~\eqref{1_20}-\eqref{1_40} for the unknown functions $f, w$, and $\phi$ will be solved numerically subject to the  boundary conditions~\eqref{3_A_10} and~\eqref{3_A_20}. Keeping in mind that we will seek solutions which are symmetrical about the plane $z=0$, numerical computations will be carried out only in the region $z>0$. In doing so,  for numerical calculations, it is convenient to introduce new compactified coordinates
\begin{equation}
	\bar x=\frac{x}{1+x},\quad \bar y=\frac{y}{1+y},
	\label{comp_coord}
\end{equation}
the use of which permits one to map the infinite region $[0,\infty)$ to the finite interval $[0,1]$.

All results of numerical calculations given below have been obtained using the package FIDISOL~\cite{fidisol} where the numerical method based on the Newton-Raphson method is employed.  This method provides  an iterative procedure for obtaining an exact solution starting from an approximate solution (an initial guess). As such initial guess, we use the solution for the infinite tube obtained in Sec.~\ref{inf_tube}. Solutions are sought on a grid of $51\times 51$ points, covering  the region of integration $0\leq \bar x, \bar y \leq 1$ given by the compactified coordinates from Eq.~\eqref{comp_coord}. This enables us to get solutions with typical relative errors on the order of $10^{-4}$.

\subsection{{\it P}-ball with a nonzero total angular momentum}
\label{non_zero_ang_mom}

Consider first the case of the type A configurations. In this case the component
$E^1_z\left( \rho, z=0\right)  = 0$  [see Eq.~\eqref{fields_5}]; this implies that there is no flux of electric field through the plane $z=0$. However, it turns out that the color electric and magnetic fields are arranged so that there is a nonzero Poynting vector ensuring the presence of
nonvanishing total angular momentum for the {\it P}-ball under consideration.

\begin{figure}[t]
\includegraphics[width=1.\linewidth]{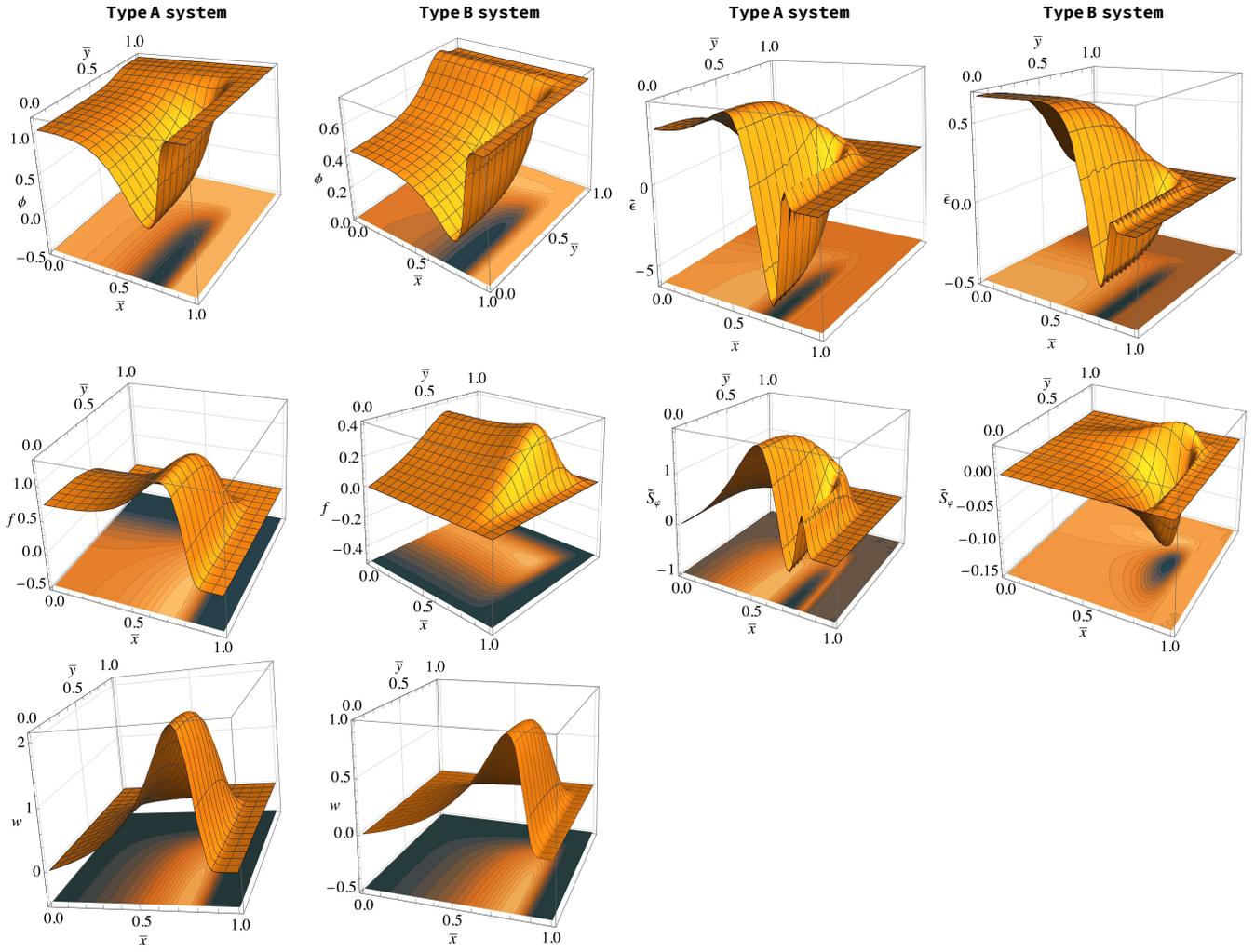}
\vspace{-0.3cm}
\caption{Typical distributions of the fields $\phi$, $f$, $w$, the total energy density of the system~$\tilde\varepsilon$ from Eq.~\eqref{2_80}, and the component $\tilde S_\varphi$ of the Poynting vector given by Eq.~\eqref{UmPoynt}. The graphs for the type A system ({\it P}-ball with a nonzero total angular momentum) are plotted by choosing the parameters  $v=1.65$, $\Lambda=0.4$, and $\alpha=1.1$;  for the type B system ({\it P}-ball with a nonzero flux of electric field through the plane $z=0$)~--~by choosing $v=1.0$, $\Lambda=0.4$, and $\alpha=0.8$. In all cases the charge/current densities are taken to be zero.
}
\label{fig_plots_fields_distr}
\end{figure}

\begin{figure}[h!]
\includegraphics[width=.9\linewidth]{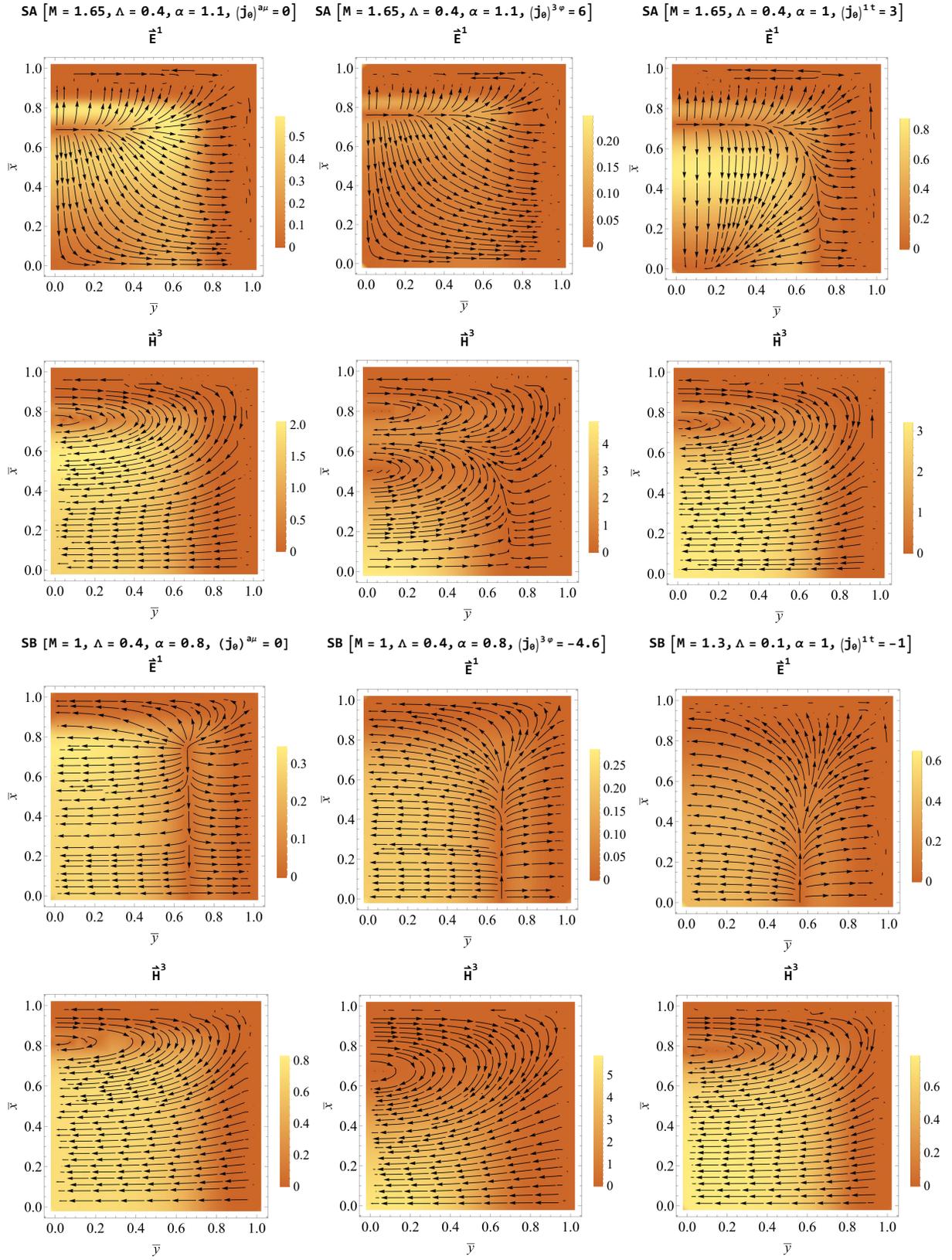}
\vspace{-0.3cm}
\caption{The dimensionless color electric, $\vec E^1$, and magnetic, $\vec H^3$, fields strength distributions for different values of the parameters for the type A systems~(SA) and type B systems~(SB). For the systems with the charge/current, the values of the free parameters $l$ and $R_0$
appearing in Eqs.~\eqref{1_300} and~\eqref{1_310} are taken to be $l=0.5$ and $R_0=1$.
}
\label{fig_elec_magn_fields}
\end{figure}

The corresponding results of computations are given in Figs.~\ref{fig_plots_fields_distr} and~\ref{fig_elec_magn_fields}. As in the case of an infinite tube of Sec.~\ref{inf_tube}, the input parameters are eigenparameters $\alpha, \Lambda$, and $v$, whose magnitudes determine  the solution completely. (Notice here that by imposing appropriate boundary conditions to Eqs.~\eqref{1_20}-\eqref{1_40}, one can reproduce the solutions for an infinite flux tube
obtained in  Sec.~\ref{inf_tube}.) Fig.~\ref{fig_plots_fields_distr} shows the typical profiles of the components of the vector potential $f$ and  $w$ and the Higgs scalar field $\phi$,  as well as of the Poynting vector and the energy density. Figure~\ref{fig_elec_magn_fields} shows the distributions of the electric, $\vec E^1$, and magnetic, $\vec H^3$, fields both in the absence of the sources in
Eqs.~\eqref{1_20} and \eqref{1_30} and when the charges/currents given in the form~\eqref{1_300} are present.

\begin{figure}[t]
\begin{minipage}[t]{.49\linewidth}
\begin{center}
	\includegraphics[width=1.\linewidth]{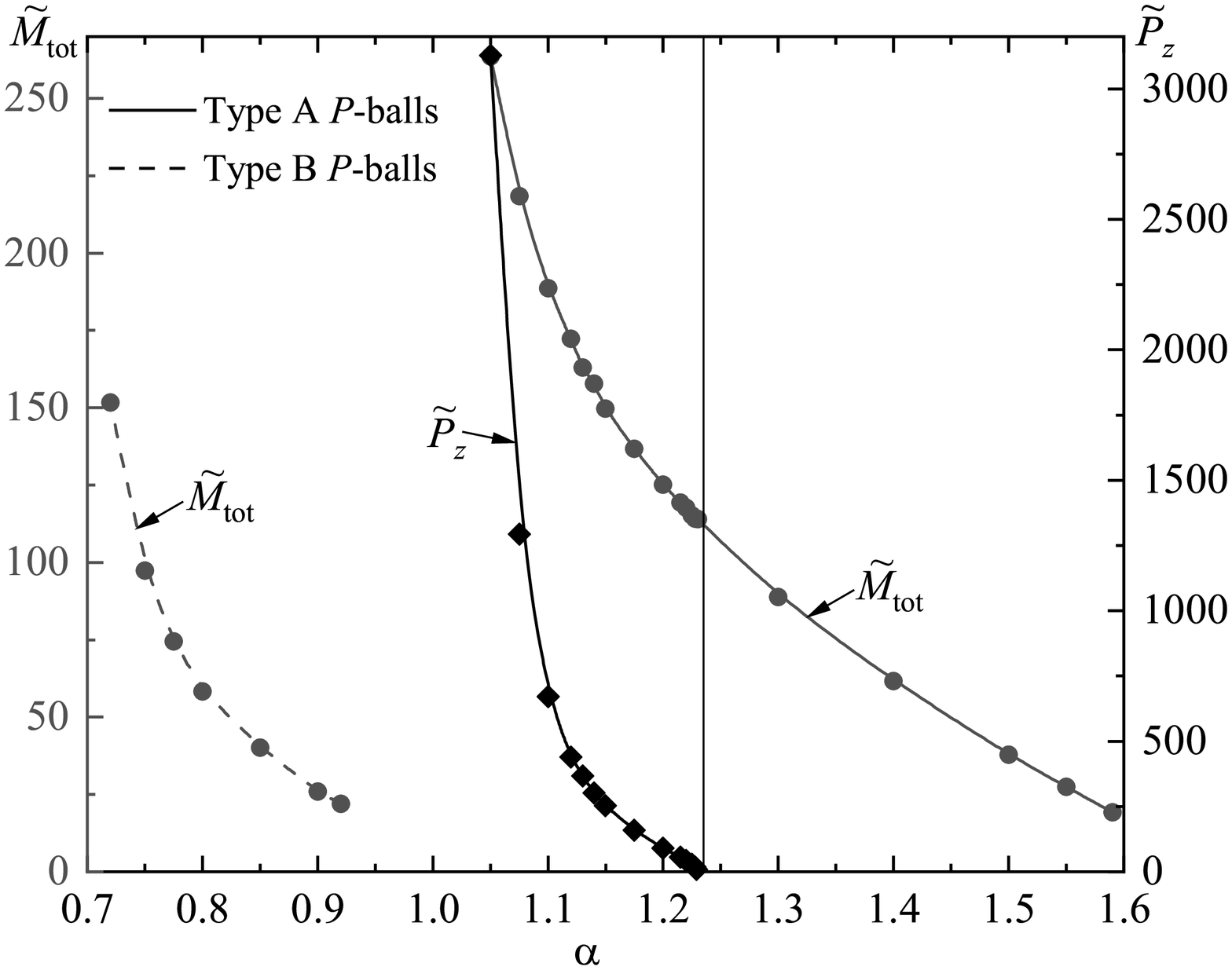}
\end{center}
\end{minipage}\hfill
\begin{minipage}[t]{.49\linewidth}
\begin{center}
	\includegraphics[width=.94\linewidth]{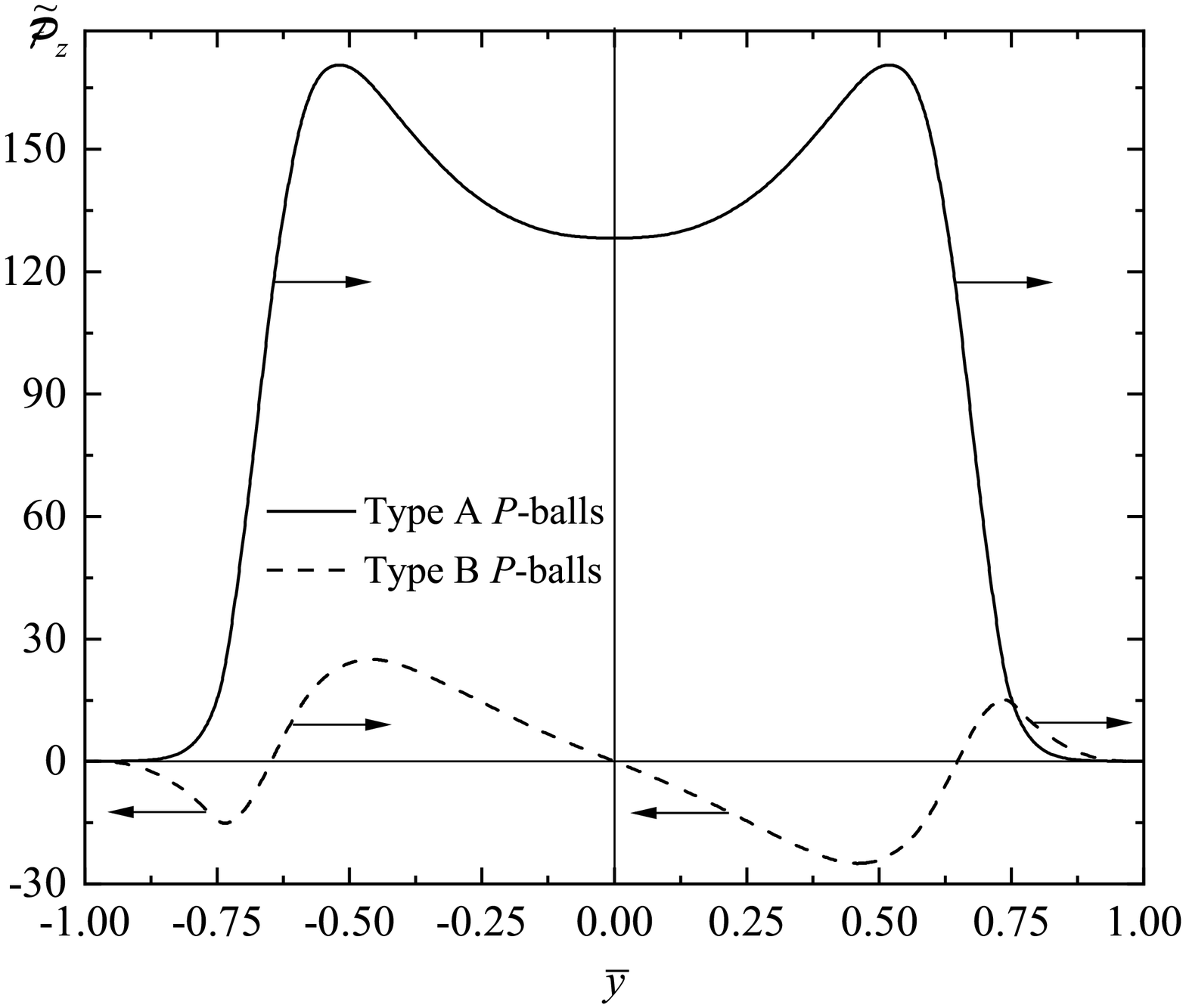}
\end{center}
\end{minipage}
\caption{Left panel: The dependence of the total mass \eqref{M_tot} and of the total angular momentum~\eqref{5_20} of the configurations under consideration on $\alpha$ for fixed values of the parameters  $v=1.65$ and $\Lambda=0.4$ (for the type A systems) and $v=1.0$ and $\Lambda=0.4$ (for the type B systems). For the type A systems, the thin vertical line separates the systems containing both the electric and magnetic fields (located to the left of the line) from those possessing only the magnetic field (located to the right of the line). Right panel: The distributions of the linear angular momentum density~\eqref{5_25} along the tube axis for the solutions shown in Fig.~\ref{fig_plots_fields_distr}. The arrows show the directions of the angular momentum density.
}
\label{fig_M_tot_P_z}
\end{figure}

Notice here the following interesting and important point. As the calculations indicate, in the absence of the charges/currents in the right-hand sides of Eqs.~\eqref{1_20} and \eqref{1_30}, regular solutions exist only for $\alpha\neq 1$. This corresponds to the fact that in this case the masses of the Proca fields $\{f,h\}$ and $\{w,k\}$ are not equal to each other. However, it is of physical interest to find solutions with $\mu_f=\mu_w$, i.e., when $\alpha= 1$; this would correspond to the fact that the Proca mass tensor
$
	\left( \mu^2 \right)^{a b, \mu}_{\phantom{a b,}\nu}
$ appearing in Eq.~\eqref{0_10} would become a scalar. It turns out that this can be achieved when the system contains a nonzero charge density $ j^{1 t}$ given by the model~\eqref{1_300}. Examples of the corresponding distributions of the electric and magnetic fields are given in the third column of Fig.~\ref{fig_elec_magn_fields}. In turn, this enables us to assume that in considering more realistic models where quarks are self-consistently described by involving spinor fields into the system, solutions with $\alpha = 1$ will also exist.

For the type A systems, the $\varphi$-component of the Poynting vector~\eqref{UmPoynt} is an even function of $z$. Since this vector is proportional to the momentum density, this ensures the presence in the system of a nonzero angular momentum density directed along the tube axis:	$\mathcal{M}_z = \rho S_\varphi$. This situation is similar to what we have for the case of infinite tubes considered in Sec.~\ref{inf_tube}, where there exists a nonzero linear angular momentum density determined by
the $\varphi$-component of the Poynting vector~\eqref{UmPoynt_inf}. But unlike the infinite tube that has an infinite total angular momentum, the finite tube will have a finite total angular momentum. In the latter case, since the function $S_\varphi$ is an even function of~$z$, for the {\it P}-ball under consideration, the total angular momentum calculated by integration over the whole volume
 $dV=\rho\, d\rho\, d\varphi\, dz$ with $0 \leq \rho < \infty$, $0 \leq \varphi \leq 2\pi$, and
 $-\infty < z < \infty$ is
\begin{equation}
	\tilde{P}_z	\equiv
	 g^2 P_z =4\pi\int\limits_{0}^{\infty} dy \int \limits_{0}^{\infty}  \tilde S_\varphi x^2 dx
\label{5_20}
\end{equation}
already written in the dimensionless form.

In turn, the total dimensionless energy density of the system under consideration can be obtained from Eq.~\eqref{0_40}, using Eqs.~\eqref{1_10}-\eqref{fields_10}, in the form
\begin{equation}
	\tilde\varepsilon \equiv \frac{g^2}{\mu_f^4}\varepsilon =
	f_{, x}^2 + f_{, y}^2 + w_{, x}^2 + w_{, y}^2 + \phi_{, x}^2
	+ \phi_{, y}^2 + 2\frac{w w_{, x}}{x} + 2\left(f^2 + w^2\right)\phi^2
	- f^2 - \alpha^2 w^2 + \frac{w^2}{x^2} + \frac{\Lambda}{4}\left(2\phi^2 - v^2\right)^2 .
\label{2_80}
\end{equation}
As is seen from Fig.~\ref{fig_plots_fields_distr}, the energy density can be negative in some regions. To see whether this will give a negative total energy of the system or not, we calculate the dimensionless total mass (energy) of the configurations under consideration,
\begin{equation}
	\tilde M_{\text{tot}}\equiv \frac{g^2}{\mu_f} M_{\text{tot}} =
	4\pi \int\limits_{0}^{\infty} dy \int \limits_{0}^{\infty} \tilde\varepsilon x dx.
\label{M_tot}
\end{equation}

The results of calculations for the total mass and total angular momentum as functions of the parameter $\alpha$ (for fixed values of two other free system parameters $v$ and $\Lambda$) are given in the left panel of Fig.~\ref{fig_M_tot_P_z} for the values of $\alpha$ for which we have succeeded in obtaining numerical solutions to the required accuracy. As the computations for the type~A systems indicate, with increasing $\alpha$, there is some critical $\alpha_{(\text{crit})}\approx 1.23$ for which the function $f\to 0$ (the electric field is switched off), and with further increase of $\alpha$, it always remains equal to zero. Consequently, the configurations with $\alpha \lesssim \alpha_{(\text{crit})}$ possess both the magnetic and electric fields, but for $\alpha \gtrsim \alpha_{(\text{crit})}$ only the magnetic field is present. In the latter case, the system under investigation becomes a configuration that can already be described within U(1) Proca scalar electrodynamics.

Apart from the total angular momentum $P_z$, it is also of interest to follow the behavior of a linear angular momentum density
\begin{equation}
	\tilde{\mathcal{P}}_z	\equiv
	 \frac{g^2}{\mu_f} \mathcal{P}_z =2\pi\int \limits_{0}^{\infty}  \tilde S_\varphi x^2 dx ,
\label{5_25}
\end{equation}
which describes a distribution of the angular momentum density along the tube axis. Examples of the corresponding distributions are given in the right panel of Fig.~\ref{fig_M_tot_P_z}. As is seen from this figure, the angular momentum for the type A systems is always directed in one direction.

Summarizing the results for the solutions of this sort, in SU(2) Higgs-Proca theory where the gauge symmetry is explicitly broken and which contains a triplet of real Higgs scalar fields, there exist particlelike solutions ({\it P}-balls) possessing a nonvanishing total angular momentum. The existence of such solutions is a consequence of the non-Abelianity of the fields. Both in pure SU(2) Proca theory (i.e., when scalar fields are absent) and in a system containing only Higgs fields (i.e., in the absence of Proca fields) there are no such solutions.

\subsection{Comparison with spinning electroweak sphalerons}
\label{comparison}

It is of interest to compare the configurations from the previous subsection with spinning electroweak sphalerons found in Refs.~\cite{Radu:2008ta,Kleihaus:2008cv}.
Those systems, obtained within $\text{SU(2)}\times \text{U(1)}$ theory involving Higgs scalar fields and a U(1) gauge field,
also possess a finite energy and nonzero total angular momentum.
The corresponding Lagrangian was taken to be
\begin{equation}
	L = - \frac{1}{4 g^2} F^a_{\mu \nu} F^{a \mu \nu}
	- \frac{1}{4 {g^\prime}^2} Y_{\mu \nu} Y^{\mu \nu}
	+ \left( D_\mu \Phi \right)^\dagger D^\mu \Phi
	- \frac{\beta}{8} \left( \Phi^\dagger \Phi - 1 	\right)^2,
\end{equation}
where
$Y_{\mu \nu}  =  \partial_\mu Y_\nu - \partial_\nu Y_\mu$ is the field tensor for the
 U(1) gauge field $Y_\mu$ and $\Phi$ is a doublet of complex Higgs fields with
$
D_\mu \Phi = \left( \partial_\mu - \frac{i}{2} Y_\mu - \frac{i}{2} \sigma^a A^a_\mu \right) \Phi
$, where $\sigma^a$ are the Pauli matrices; $g$ and $g^\prime$ are the gauge coupling constants and $\beta$ is the self-coupling constant for the Higgs field.

The presence of the U(1) gauge field ensures the following relation between the total angular momentum $ P_z$ and electric charge $Q$~\cite{Radu:2008ta}
\begin{equation}
	 P_z = \frac{n}{g g^\prime} Q ,
\end{equation}
where $n$ is an integer. Consequently, it is seen that the presence of the angular momentum in $\text{SU(2)}\times \text{U(1)}$ theory is directly related to the presence of
a U(1) electric charge.

Thus, there is a fundamental difference in the manner of the occurrence of the angular momentum in the systems of
Refs.~\cite{Radu:2008ta,Kleihaus:2008cv} and those considered in the present study.
This, in turn, means that the conditions imposed by the no-go theorem of Ref.~\cite{Volkov:2003ew}, according to which there are no stationary and axially symmetric spinning excitations for the 't~Hooft-Polyakov monopoles,
Julia-Zee dyons, sphalerons, and also vortices, can be circumvented either by introducing an extra U(1) gauge field (as is done in Refs.~\cite{Radu:2008ta,Kleihaus:2008cv})
or by explicitly violating the gauge symmetry, as is done in the present paper when there is no need to introduce an extra U(1) gauge field.

\subsection{{\it P}-ball with a nonzero flux of electric field
}
\label{el_flux}

In this subsection we consider the case of the type B configurations, which already have $E_z(\rho, z=0) \neq 0$, i.e., there is a flux of electric field through the plane  $z=0$. For this case, Fig.~\ref{fig_plots_fields_distr} shows the typical profiles of the components of the vector potential $f$ and $w$, the Higgs scalar field $\phi$, and also the Poynting vector and the energy density. In turn, Fig.~\ref{fig_elec_magn_fields} shows the distributions of the electric, $\vec E^1$, and magnetic, $\vec H^3$, fields both in the absence of the sources in Eqs.~\eqref{1_20} and \eqref{1_30} and when the charges/currents given in the form~\eqref{1_310} are present.

As in the case of the type A configurations considered in the previous subsection, for the type B systems,
it is possible to obtain regular solutions with equal masses of the Proca fields $\{f,h\}$ and $\{w,k\}$ only in the presence of a nonzero charge density $ j^{1 t}$. Its choice in the form~\eqref{1_310} enables us to obtain the required solutions with typical distributions of the electric and magnetic fields shown in the third column of Fig.~\ref{fig_elec_magn_fields}. Since in this case the component $S_\varphi$ of the Poynting vector is an odd function of $z$, so that, despite the presence of a nonzero angular momentum in the hemispace $z>0$ (see Fig.~\ref{fig_plots_fields_distr}), the total angular momentum calculated over the whole space will be zero, that is, $P_z = 0$.

In turn, the left panel of Fig.~\ref{fig_M_tot_P_z} shows the dependence of the total mass of the type B configurations on the parameter~$\alpha$. Unlike the type A systems, in this case, for all values of $\alpha$ for which we were able to obtain solutions to the required accuracy, the systems always contain both the electric and magnetic fields.

As regards the linear angular momentum~\eqref{5_25}, unlike the type A systems in which the angular momentum is always directed in one direction, for the type B configurations, the direction of the angular momentum varies along the tube axis, as is depicted in the right panel of Fig.~\ref{fig_M_tot_P_z}.

Summarizing the results of this subsection, it is shown that there exist {\it P}-ball type solutions possessing a flux of electric field through the plane of symmetry  $z=0$. For such configurations, to the right and to the left of this plane, there are equal in modulus but oppositely directed angular momenta;
consequently, the total angular momentum of the system vanishes.

\section{Discussion and conclusions}
\label{concl}

In the present study, we have shown that in SU(2) Higgs-Proca theory there exist regular solitonlike static solutions possessing a finite total energy. Such solutions can be used  for the description of localized Proca-ball-type configurations. The latter are in equilibrium due to the force balance: the non-Abelian Proca field is purely repulsive, whereas the Higgs scalar field is purely attractive. The crucial feature of such systems is that they have the following physically interesting properties: they either possess a nonzero total angular momentum (despite the fact that the solutions are static) or have a flux of electric field through the plane of symmetry  $z=0$. It is important to notice that separately neither in SU(2) Proca theory nor in the system supported only by the Higgs scalar field the existence of such solutions is possible. From the mathematical point of view, that is because the non-Abelian Proca equations~\eqref{1_20} and \eqref{1_30} are Schr\"{o}dinger-type equations for the functions $f$ and $w$; for regular solutions of these equations to exist, one has to have a potential well that, in the present case, is created by the scalar field  $\phi$ obeying Eq.~\eqref{1_40}. In the absence of such a scalar field, Eqs.~\eqref{1_20} and \eqref{1_30} will have only singular solutions. In turn, according to Derrick's theorem~\cite{Derrick:1964ww}, in the absence of extra fields,  the equation for the scalar field $\phi$
 has no $(3+1)$-dimensional regular solutions.

The following feature of the systems under consideration should be mentioned. In the absence of external sources [that is, when in Eq.~\eqref{0_20} the four-current $j^{a \mu}=0$] we were able to obtain regular solutions to the set of equations~\eqref{1_20}-\eqref{1_40} only in the case where masses of the Proca fields $\{f,h\}$ and $\{w,k\}$ were different. However, it is of special physical interest to get solutions for a system in which all components of the Proca field would have equal masses. It turns out that this can be achieved if a nonzero charge density is included in the system. In our case, to model the charge, we have used toy models in the form~\eqref{1_300} and~\eqref{1_310}; this enabled us to get solutions with equal Proca-field masses. In a more realistic case, the toy models used here must be replaced by suitable four-currents created, for example, by a spinor field $\psi$ in the form
$j^{a\mu}\sim\bar\psi\gamma^\mu\sigma^a\psi$, where $\gamma^\mu$ and $\sigma^a$ are the Dirac and Pauli matrices, respectively. If it would be possible to obtain regular solutions for such a self-consistent SU(2)-Higgs-Proca-Dirac system, it might be an argument in favor of the possibility of using such systems for, e.g., the approximate modeling of the confinement phenomenon in QCD. The hope for obtaining solutions for such a system is related to the fact that we have already found particlelike solutions in SU(2) Higgs-Proca-Dirac
theory~\cite{Dzhunushaliev:2020eqa, Dzhunushaliev:2019ham}, as well as monopole solutions in SU(2) Yang-Mills-Dirac theory~\cite{Dzhunushaliev:2020qwf, Burtebayev:2021qtv}.

It is evident that the problem of the existence of Proca fields in nature is of great importance (as briefly discussed, e.g., in Refs.~\cite{Dzhunushaliev:2021oad,Dzhunushaliev:2021vwn}).
As a first possibility, one may assume that the Proca fields are fundamental; but this implies the violation of the gauge principle according to which all fundamental integer-spin fields must be gauge invariant.
The second possibility presupposes that the Proca fields are phenomenological; this means that they are introduced in a theory to give an approximate description of some nonperturbative quantum phenomena. In any case, there are some motivations for introducing such fields, as discussed, in particular, in the literature devoted to various aspects of using Proca
fields~\cite{Brito:2015pxa,Herdeiro:2017fhv,Dzhunushaliev:2019kiy,Herdeiro:2019mbz,Dzhunushaliev:2019uft,Bustillo:2020syj,Heisenberg:2017xda,Kase:2018voo,Arkani-Hamed:2008hhe,Pospelov:2008jd,DeFelice:2016yws,DeFelice:2016uil,Tu:2005ge,Dzhunushaliev:2019sxk,Dzhunushaliev:2020eqa,Dzhunushaliev:2021uit}.

Let us also note the following properties of the solutions obtained:
\begin{itemize}
\item[(a)] Regular finite-size tube solutions exist both with and without external sources (charges/currents).
\item[(b)] According to the no-go theorem of Ref.~\cite{Volkov:2003ew},  there are no stationary and axially symmetric spinning excitations for all known topological solitons in the one-soliton sector, that is, for the 't~Hooft-Polyakov monopoles, Julia-Zee dyons, sphalerons, and also vortices. The existence of
the solutions with a nonvanishing total angular momentum obtained here is possible since we
circumvent the conditions of this theorem in view of the fact that (i)~the gauge symmetry is explicitly broken by introducing the mass tensor for the non-Abelian field,
and  (ii)~the solutions at hand are topologically trivial.
Notice also that the ways to circumvent the conditions imposed by this no-go theorem
employed in Refs.~\cite{Radu:2008ta,Kleihaus:2008cv} and in the present study are fundamentally different;
in Refs.~\cite{Radu:2008ta,Kleihaus:2008cv},  an extra U(1) gauge field is introduced, whereas in the present work the gauge symmetry is explicitly broken.
As a result, for the configurations of Refs.~\cite{Radu:2008ta,Kleihaus:2008cv}, there is some relation between an angular momentum and U(1) electric charge (and no charge~= no angular momentum),
whereas in our case there is no such a relation.
\item[(c)] There is a profound difference between {\it Q}-balls and {\it P}-balls related to the fact that the angular momentum of a {\it Q}-ball is provided by a spinning scalar field, whereas the angular momentum of a {\it P}-ball is created by static crossed electric and magnetic Proca fields. In a certain sense, the systems with a nonzero total angular momentum, in analogy to Wheeler’s conception of ``charge without charge'', may be called the configurations possessing ``rotation without rotation''.
\item[(d)] The inclusion in the system of a charge permits one to obtain solutions with equal masses of all components of the vector potential. In this case the Proca field mass tensor becomes a scalar quantity; this enables one to avoid ambiguities in defining the energy-momentum tensor [see the discussion after Eq.~\eqref{emt_gen}].
\item[(e)] In the general case the {\it P}-balls  are essentially non-Abelian objects. However, it is demonstrated that for the configurations with a nonzero total angular momentum there exists some critical value of the system parameter~$\alpha$  for which the electric field is switched off and only the magnetic field remains nonzero. In this case the system becomes a configuration that can already be described within U(1) Proca scalar electrodynamics.
\item[(f)]  For the systems obtained the Meissner-like effect is seen; the electric, magnetic, and scalar fields expel one another. In a certain sense, this is analogous to the well-known effect in dual QCD when a color electric field is expelled by magnetic monopoles.
\item[(g)] There are {\it P}-ball solutions supported only by the non-Abelian Proca and scalar fields, without involving any sources of such fields (such as quarks, for example).
This can be regarded as a close analogue of the problem of obtaining glueballs in QCD; this enables us to assume that {\it P}-balls may play the role of glueballs in SU(2) Higgs-Proca theory.
\end{itemize}

The solutions obtained have no nodes and can be regarded as fundamental {\it P}-balls
(cf. the fundamental {\it Q}-balls of Refs.~\cite{Volkov:2002aj,Kleihaus:2005me}). For such configurations, for fixed values of the system parameters $\alpha, \Lambda$, and $v$, there are the only values of the total energy/mass and angular momentum. Apparently, for the same values of the above parameters, solutions with nodes (if they exist) will describe excited states of the system whose
total energy/mass and angular momentum will already be different.

In conclusion, a few words should be said about the question of stability of the systems under investigation. The configurations considered here are described by nontopological solutions (no topological charge) and supported by real fields (no conserved charge or particle number).
For this reason, it is impossible to apply to them the stability criteria based on the conservation of the topological charge (as is done, for example, in the case of topological (kinklike) solutions for scalar fields or in the case of the 't~Hooft-Polyakov monopole solution~\cite{Rajaraman:1982is}) or related to the presence of some other conserved quantities (for instance, the isospin in SU(2) theory with a Higgs-type field~\cite{Friedberg:1976az} or a nonzero charge associated with a complex scalar field~\cite{Friedberg:1976me}).

In this connection we see two possible ways of studying the stability of the systems under consideration.
First, one can examine the stability with respect to axisymmetric perturbations (both linear and nonlinear), as is done for nontopological  systems (see, e.g., Ref.~\cite{Lee:1991ax}) and in the case of various objects supported by non-Abelian fields (see, e.g., the problems with linear~\cite{Straumann:1989tf} and nonlinear~\cite{Zhou:1991nu} perturbations).
Second, it is possible to study the stability within catastrophe theory~\cite{Kusmartsev:1990cr}. In both cases, for axially symmetric systems of the type considered in the present paper, this is a technically complicated problem that requires a careful analysis, and we plan to do this in a separate work.

\section*{Acknowledgments}

This research has been funded by the Science Committee of the Ministry of Education and Science of the Republic of Kazakhstan (Grant No.~BR10965191 ``Complex research in nuclear and radiation physics, high-energy physics and cosmology for development of the competitive technologies''). We are also grateful to the Research Group Linkage Programme of the Alexander von Humboldt Foundation for the support of this research.


\begin{thebibliography}{99}

\bibitem{Volkov:2002aj}
M.~S.~Volkov and E.~Wohnert,
Spinning Q-balls,
Phys. Rev. D \textbf{66}, 085003 (2002).

\bibitem{Kleihaus:2005me}
B.~Kleihaus, J.~Kunz, and M.~List,
Rotating boson stars and Q-balls,
Phys. Rev. D \textbf{72}, 064002 (2005).

\bibitem{Volkov:2003ew}
M.~S.~Volkov and E.~Wohnert,
On the existence of spinning solitons in gauge field theory,
Phys. Rev. D \textbf{67}, 105006 (2003).

\bibitem{Radu:2008ta}
E.~Radu and M.~S.~Volkov,
Spinning electroweak sphalerons,
Phys. Rev. D \textbf{79}, 065021 (2009).

\bibitem{Kleihaus:2008cv}
B.~Kleihaus, J.~Kunz, and M.~Leissner,
Electroweak sphalerons with spin and charge,
Phys. Lett. B \textbf{678}, 313 (2009).

\bibitem{Shuryak2021}
 E.~Shuryak, {\it Nonperturbative topological phenomena in QCD and related theories} (Springer Nature Switzerland AG.,
Cham, Switzerland, 2021).

\bibitem{Brito:2015pxa}
  R.~Brito, V.~Cardoso, C.~A.~R.~Herdeiro, and E.~Radu,
Proca stars: Gravitating Bose-Einstein condensates of massive spin 1 particles,
  Phys.\ Lett.\ B {\bf 752}, 291 (2016).

\bibitem{Herdeiro:2017fhv}
  C.~A.~R.~Herdeiro, A.~M.~Pombo, and E.~Radu,
  Asymptotically flat scalar, Dirac and Proca stars: discrete vs. continuous families of solutions,
  Phys.\ Lett.\ B {\bf 773}, 654 (2017).

\bibitem{Dzhunushaliev:2019kiy}
  V.~Dzhunushaliev and V.~Folomeev,
  Dirac star in the presence of Maxwell and Proca fields,
  Phys.\ Rev.\ D {\bf 99},  104066 (2019).

\bibitem{Herdeiro:2019mbz}
   C.~Herdeiro, I.~Perapechka, E.~Radu, and Y.~Shnir,
  Asymptotically flat spinning scalar, Dirac and Proca stars,
  Phys.\ Lett.\ B {\bf 797}, 134845 (2019).

\bibitem{Dzhunushaliev:2019uft}
V.~Dzhunushaliev and V.~Folomeev,
Dirac star with SU(2) Yang-Mills and Proca fields,
Phys. Rev. D \textbf{101},  024023 (2020).

\bibitem{Bustillo:2020syj}
J.~C.~Bustillo, N.~Sanchis-Gual, A.~Torres-Forn\'e, J.~A.~Font, A.~Vajpeyi, R.~Smith, C.~Herdeiro, E.~Radu, and S.~H.~W.~Leong,
GW190521 as a Merger of Proca Stars: A Potential New Vector Boson of $8.7\times 10^{-13}$  eV,
Phys. Rev. Lett. \textbf{126},  081101 (2021).

\bibitem{Heisenberg:2017xda}
L.~Heisenberg, R.~Kase, M.~Minamitsuji, and S.~Tsujikawa,
Hairy black-hole solutions in generalized Proca theories,
Phys. Rev. D \textbf{96},  084049 (2017).

\bibitem{Kase:2018voo}
R.~Kase, M.~Minamitsuji, S.~Tsujikawa, and Y.~L.~Zhang,
Black hole perturbations in vector-tensor theories: The odd-mode analysis,
J. Cosmol. Astropart. Phys. 02 (2018) 048.

\bibitem{Arkani-Hamed:2008hhe}
N.~Arkani-Hamed, D.~P.~Finkbeiner, T.~R.~Slatyer, and N.~Weiner,
A theory of dark matter,
Phys. Rev. D \textbf{79}, 015014 (2009).

\bibitem{Pospelov:2008jd}
  M.~Pospelov and A.~Ritz,
  Astrophysical signatures of secluded dark matter,
  Phys.\ Lett.\ B {\bf 671}, 391 (2009).

\bibitem{DeFelice:2016yws}
A.~De Felice, L.~Heisenberg, R.~Kase, S.~Mukohyama, S.~Tsujikawa, and Y.~l.~Zhang,
Cosmology in generalized Proca theories,
J. Cosmol. Astropart. Phys. 06 (2016) 048.

\bibitem{DeFelice:2016uil}
A.~De Felice, L.~Heisenberg, R.~Kase, S.~Mukohyama, S.~Tsujikawa, and Y.~l.~Zhang,
Effective gravitational couplings for cosmological perturbations in generalized Proca theories,
Phys. Rev. D \textbf{94}, 044024 (2016).

\bibitem{Tu:2005ge}
  L.~C.~Tu, J.~Luo, and G.~T.~Gillies,
  The mass of the photon,
  Rep.\ Prog.\ Phys.\  {\bf 68}, 77 (2005).

\bibitem{Dzhunushaliev:2019sxk}
V.~Dzhunushaliev and V.~Folomeev,
Proca tubes with the flux of the longitudinal chromoelectric field and the energy flux/momentum density,
Eur. Phys. J. C \textbf{80},  1043 (2020).

\bibitem{Dzhunushaliev:2020eqa}
V.~Dzhunushaliev, V.~Folomeev, T.~Kozhamkulov, A.~Makhmudov, and T.~Ramazanov,
Non-Abelian Proca theories with extra fields: particlelike and flux tube solutions,
Phys. Scr. \textbf{95},  074013 (2020).

\bibitem{Dzhunushaliev:2021uit}
V.~Dzhunushaliev, V.~Folomeev, and A.~Tlemisov,
Linear energy density and the flux of an electric field in Proca tubes,
Symmetry \textbf{13}, 640 (2021).

\bibitem{Dzhunushaliev:2021oad}
V.~Dzhunushaliev and V.~Folomeev,
Axially symmetric particlelike solutions with the flux of a magnetic field in the non-Abelian Proca-Higgs theory,
Phys. Rev. D \textbf{104}, 116027 (2021).

\bibitem{Dzhunushaliev:2021vwn}
V.~Dzhunushaliev and V.~Folomeev,
Axially symmetric Proca-Higgs boson stars,
Phys. Rev. D \textbf{104},  104024 (2021).

\bibitem{fidisol}
W. Sch\"{o}nauer and R. Wei{\ss},
Efficient vectorizable PDE solvers,
J. Comput. Appl. Math. {\bf 27}, 279 (1989).

\bibitem{Derrick:1964ww}
G.~H.~Derrick,
Comments on nonlinear wave equations as models for elementary particles,
J. Math. Phys. (N.Y.) \textbf{5}, 1252 (1964).

\bibitem{Dzhunushaliev:2019ham}
V.~Dzhunushaliev, V.~Folomeev, and A.~Makhmudov,
Non-Abelian Proca-Dirac-Higgs theory: Particlelike solutions and their energy spectrum,
Phys. Rev. D \textbf{99},  076009 (2019).

\bibitem{Dzhunushaliev:2020qwf}
V.~Dzhunushaliev, V.~Folomeev, and A.~Serikbolova,
Monopole solutions in SU(2) Yang-Mills-plus-massive-nonlinear-spinor-field theory,
Phys. Lett. B \textbf{806}, 135480 (2020).

\bibitem{Burtebayev:2021qtv}
N.~Burtebayev, V.~Dzhunushaliev, V.~N.~Folomeev, J.~Kunz, A.~Serikbolova, and A.~Tlemisov,
Mass gap for a monopole interacting with a nonlinear spinor field,
Phys. Rev. D \textbf{104},  056010 (2021).

\bibitem{Rajaraman:1982is}
  R.~Rajaraman,
  {\it An Introduction to Solitons and Instantons in Quantum Field Theory}
 (North-Holland, Amsterdam,  1982).

\bibitem{Friedberg:1976az}
R.~Friedberg, T.~D.~Lee, and A.~Sirlin,
Gauge field nontopological solitons in three space dimensions. 1.,
Nucl. Phys.  \textbf{B115}, 1 (1976).

\bibitem{Friedberg:1976me}
R.~Friedberg, T.~D.~Lee, and A.~Sirlin,
A class of scalar-field soliton solutions in three space dimensions,
Phys. Rev. D \textbf{13}, 2739 (1976).

\bibitem{Lee:1991ax}
T.~D.~Lee and Y.~Pang,
Nontopological solitons,
Phys. Rep. \textbf{221}, 251 (1992).

\bibitem{Straumann:1989tf}
N.~Straumann and Z.~H.~Zhou,
Instability of the Bartnik-McKinnon solution of the Einstein-Yang-Mills equations,
Phys. Lett. B \textbf{237}, 353 (1990).

\bibitem{Zhou:1991nu}
Z.~h.~Zhou and N.~Straumann,
Nonlinear perturbations of Einstein-Yang-Mills solitons and non-Abelian black holes,
Nucl. Phys. \textbf{B360}, 180 (1991).

\bibitem{Kusmartsev:1990cr}
F.~V.~Kusmartsev, E.~W.~Mielke, and F.~E.~Schunck,
Gravitational stability of boson stars,
Phys. Rev. D \textbf{43}, 3895 (1991).

\end{thebibliography}
\end{document}